# INDOOR AND OUTDOOR *IN SITU* HIGH-RESOLUTION GAMMA RADIATION MEASUREMENTS IN URBAN AREAS OF CYPRUS

E. Svoukis and H. Tsertos[*]

*Department of Physics, University of Cyprus, Nicosia, Cyprus*

**Abstract**

*In situ,* high-resolution, gamma-ray spectrometry of a total number of 70 outdoor and 20 indoor representative measurements were performed in preselected, common locations of the main urban areas of Cyprus. Specific activities and gamma absorbed dose rates in air due to the naturally occurring radionuclides of $^{232}$Th and $^{238}$U series, and $^{40}$K are determined and discussed. Effective dose rate to the Cyprus population due to terrestrial gamma radiation is derived directly from this work. The results obtained outdoors match very well with those derived previously by high-resolution gamma spectrometry of soil samples, which were collected from the main island bedrock surface. This implies that the construction and building materials in urban areas do not affect the external gamma dose rate; thus they are mostly of local origin. Finally, the indoor/outdoor gamma dose ratio was found to be $1.4 \pm 0.5$.

---

[*] **Corresponding author.** E-mail address: *tsertos@ucy.ac.cy*, Fax: +357-22892821.

# INTRODUCTION

One of the main external sources of irradiation to the human body is represented by the gamma radiation (terrestrial environmental background radiation) emitted by naturally occurring radioisotopes. The most prominent naturally occurring radioisotopes are $^{40}$K and the radionuclides from the $^{232}$Th and $^{238}$U series with their decay products, which exist at trace levels in all ground formations. During the last decades, extensive surveys have been carried out worldwide to determine activity concentration levels and associated dose rates due to terrestrial gamma radiation (see UNSCEAR 2000 report [1], and further references cited therein). Such investigations are important not only for assessing population exposure and performing epidemiological studies, but also for serving as a reference (baseline data) to possible environmental contaminations due to human activities.

The first systematic data on environmental gamma radiation and associated dose rates in Cyprus were obtained by a series of measurements[2,3,4], which have been carried out at the Nuclear Physics Laboratory of the Department of Physics, University of Cyprus, in the framework of a project called "radioisotopes". The standard soil sampling method and the laboratory measurement technique by means of high-resolution gamma spectrometry were employed for this work. The first results were based on 28 samples collected from the main geological rock types in the island[2]. An extensive survey of terrestrial gamma radioactivity was performed after the systematic collection of 115 soil samples, stemming from the main island bedrock surface[3,4]. However, the soil sampling method cannot be applied to urban areas, where most of the Cyprus population is, since building and other construction materials might affect the external dose rate. Moreover, this type of study cannot be used, at all, to determine directly the gamma dose rate in the interior of buildings and dwellings (indoor environments), where the people spend most of their time (about 80%). The *in situ* high-resolution gamma spectrometry is highly requested for reliable results in indoor/outdoor urban environments[5].

The main goal of the present study is to provide the first *in situ* high-resolution gamma spectrometry measurements outdoors and indoors in the main urban areas of Cyprus, and to determine accurately the effective dose to the population due to terrestrial gamma radiation. The *in situ* gamma spectrometry was performed, utilizing a high-resolution portable Ge detector.

## MATERIALS AND METHODS

### Gamma-ray spectroscopic system

*In situ* gamma measurements were conducted using a portable high-purity germanium (HPGe) detector (coaxial cylinder of 50.1 *mm* in diameter and 44 *mm* in length) with a relative efficiency of *15%* and an energy resolution (FWHM) of 1.8 *keV* at the 1.33 *MeV* reference transition of $^{60}$Co. The detector is coupled to a cryostat, dipped into a small dewar (capacity 3 *l*) filled with liquid nitrogen, that features in all-attitude capability. The whole system was mounted on a movable small vehicle at a fixed position, with the (detector) Ge crystal facing the ground at a height of 1 *m*. For this survey, the "ORTEC Nomad Plus Portable Spectroscopy System" was used, which incorporates a high-voltage power supply, a spectroscopy amplifier and a multi-channel buffer (MCB) consisting of a 16k ADC in a compact unit. An advanced multi-channel analyser (MCA) emulation software (GammaVision-32[6]) enabled control of the power supply and amplifier as well as control of the data acquisition, storage, display and analysis of the acquired spectra.

**Counting and data analysis**

During the period February – June 2006, a total number of 70 outdoor and 20 indoor measurements were carried out in accessible urban areas of the Republic of Cyprus, which is divided into the five main districts shown in Figure 1. Table 1 presents the district, the population[7], and the number of outdoor/indoor measurements performed in each district. The present study was concentrated on - and around - the five main towns indicated in Fig. 1, where live more than 90% of the people[7]. Outdoor measurements were taken near public buildings (hospitals, schools, civil services, etc), pavements of main city roads and frequented national parks and gardens. Indoor measurements were carried out in the interior of public buildings of the Lefkosia district. The duration of each measurement was 1 $h$ with a statistical accuracy of about 3 to 8% for the measured photopeaks of interest.

**Derivation of the specific activities from the *in situ* gamma-ray spectra**

Measured gamma radiation spectra were analysed using the ORTEC M-1 analysis code[8]. This code uses the GammaVision peak fitting and analysis algorithm combined with the "1-meter" methodology developed by the U. S**.** Department of Energy-Environmental Measurements Laboratory (EML) (see Beck *et al.*[9]), and has been further developed by Helfer and Miller[10]. This methodology was initially applied to field gamma measurements to account for directly activity concentrations due to natural or human-made radionuclides. It reduces a complex measurement problem to the product of three, simply determined factors, which have been calculated for a range of detector types and ground (soil) conditions and are tabulated within the program. Over the years, this technique has been adopted and improved by other authors; see, e.g. the ICRU report 53[11], and Clouvas et al.[12] for application of this method to measurements in an indoor environment.

According to the M-1 analysis procedure[8], the measured absorption photopeak count rate is converted into a specific activity, $A_E$ (*Bq m$^{-2}$* or *Bq g$^{-1}$*), in the ground (soil) by the product of the three factors that correspond to each of the identified radionuclides. The only necessary parameters to insert in the code are the detector efficiency, the detector aspect ratio calculated from the crystal dimensions (length/diameter), and the deposition profile parameter ($\alpha/\rho$ values)[8]. The latter is equal to 0 for uniformly distributed natural emitters.

In indoor environments, however, the situation becomes more complex: the gamma-source geometry is generally unknown, and gamma radiation comes not only from the ground, but also from the walls and roofs. The building materials act as sources of radiation and also as shields against outdoor radiation. In this case, the walls efficiently absorb the gamma rays emitted outdoors, and the indoor absorbed rate depends mainly on the activity concentrations of the natural radionuclides in the building materials[5].

Most of the buildings and houses in Cyprus contain strengthened reinforced concrete frames, walls from bricks, roofs from concrete and tiles above the reinforced concrete floor. The walls are covered by plaster. Building materials are usually of local origin and therefore radionuclide concentrations are similar to local soil. In such cases, detailed Monte-Carlo simulations that include different indoor geometries and gamma-source distributions are needed to obtain reliable results on the specific activities and associated absorbed dose rates. However, Clouvas et al.[12] have shown, by performing such calculations, that the results obtained do not depend strongly on the different parameters such as the dimension of the rooms, the thickness of walls, the density of the building materials, and the gamma-source geometry, but they mostly depend on the activity concentrations of the natural radionuclides contained in the building materials. Indeed, the build-up-factors, which enter into the determination of the gamma absorbed dose rates, do not differ by more than 10-12% between outdoor and indoor measurements[5]. Therefore, the same analysis code (ORTEC M-1) and the procedure described above were applied to determine the specific activities of the natural radionuclides in the present indoor measurements. Under these circumstances, the indoor/outdoor ratio of the absorbed dose rates in air is expected to be higher than one as a result of the change in gamma-source geometry.

The radionuclides that were considered in the present analysis are: $^{214}$Bi (main transition at ~609 *keV*, ~46.1%), $^{228}$Ac (~911 *keV*, ~29%), and $^{40}$K (~1461 *keV*, ~10.7%). The $^{232}$Th and $^{238}$U activities were estimated from the corresponding activities of the short-lived $^{228}$Ac and $^{214}$Bi radioisotopes, under the assumption that secular equilibrium was reached between $^{232}$Th and $^{238}$U and their decay products. Concerning this point, one should notice that equilibrium is common in rocks older than $10^6$ *years*, and that the $^{232}$Th series may be considered in equilibrium in most geological environments (Chiozzi et al.[13]). On the contrary, the $^{40}$K activity was determined directly from the measurement of the strong gamma-ray photopeak at 1460.75 *keV* (see ref.[2] for details).

The Minimum Detectable Activity (MDA) was calculated to be ~4 *Bq kg$^{-1}$* for $^{232}$Th, ~2 *Bq kg$^{-1}$* for $^{238}$U, and ~6 *Bq kg$^{-1}$* for $^{40}$K, for a counting time of 1 *hour*. Only naturally occurring radionuclides were detected in the measured spectra, with an exception of a small contamination of $^{137}$Cs due to the Chernobyl nuclear accident. The latter was present in some outdoor measurements with a mean activity of ~270 *Bq m$^{-2}$* or ~5.5 *Bq kg$^{-1}$*. Its contribution to the total gamma dose rate is insignificant and, therefore, it has been neglected.

**Derivation of the absorbed dose rates and effective dose rates**

If naturally occurring radioactive nuclides are uniformly distributed, the dose rates at 1 m above the ground surface, D (in units of *nGy h$^{-1}$*), can be calculated by the following equation[2,14]:

$$D = A_E C_F, \qquad (1)$$

where $A_E$ is the measured specific activity (*Bq kg$^{-1}$*) of the radionuclide and $C_F$ is a conversion factor (*nGy h$^{-1}$ per Bq kg$^{-1}$*). In the present work, mean values of $C_F$ = 0.528, 0.389 and 0.039 *nGy h$^{-1}$ per Bq kg$^{-1}$* were used in the dose rate calculations for $^{232}$Th series, $^{238}$U series, and $^{40}$K, respectively (see ref.[2] for details).

The annual effective dose rate to the population, $H_E$, is calculated by the following equation:

$$H_E = D\,T\,F, \qquad (2)$$

where $D$ is the calculated total dose rate, $T$ is the occupancy time ($T = f \times 24 \times 365.25$ *hy$^{-1}$*, $f$ is the occupancy factor with values of 0.2 and 0.8 for outdoor and indoor measurements, respectively), and $F$ is the conversion factor (0.7 *Sv Gy$^{-1}$*)[1].

# RESULTS AND DISCUSSION

In Table 2, the range and mean value of the outdoor specific activities of the naturally occurring $^{232}$Th, $^{238}$U, and $^{40}$K radionuclides for each of the 5 main urban districts are presented. The specific-activity values ranged from 1.7 to 25.0 *Bq kg$^{-1}$* for $^{232}$Th, from 5.4 to 31.8 *Bq kg$^{-1}$* for $^{238}$U, and from 81 to 388 *Bq kg$^{-1}$* for $^{40}$K. The average specific activity of $^{232}$Th, $^{238}$U, and $^{40}$K ranged from 9.9 to 11.9 *Bq kg$^{-1}$* with a mean (A.M. ± S.D.) of 10.6 ± 5.1 *Bq kg$^{-1}$*, from 11.6 to 16.5 *Bq kg$^{-1}$* with a mean of 14.2 ± 5.7 *Bq kg$^{-1}$*, and from 146 to 160 *Bq kg$^{-1}$* with a mean of 153 ± 56 *Bq kg$^{-1}$*, respectively.

The present results on the outdoor specific activity of the three naturally occurring radionuclides are in very good agreement with the corresponding values obtained previously by laboratory high-resolution gamma spectrometry of soil samples from the main geological formations of Cyprus[3,4]. The 5 districts covered by the present *in situ* measurements belong to 4 different geological formations of sedimentary origin, which have been studied and discussed in details (with respect to naturally occurring radioactivity) elsewhere[3,4]. According to these, the averaged specific activities over all main sedimentary formations were found to be (A.M ± S.D.)[3]: 9.3 ± 8.1, 13.2 ± 8.8, and 147 ± 98 *Bq kg$^{-1}$*, for $^{232}$Th, $^{238}$U and $^{40}$K, respectively. It should be pointed out here that the median values derived from population-weighted data available worldwide[1] are: 30, 35, and 400 *Bq kg$^{-1}$*, respectively. Hence, the maximum activity values obtained from the present *in situ* measurements of naturally occurring $^{232}$Th, $^{238}$U and $^{40}$K radioisotopes are still below the corresponding worldwide mean activity values reported in the UNSCEAR 2000 Report[1].

As expected, the average activity concentrations of the three radionuclides are higher in the indoor measurements: the specific-activity values ranged from 4.9 to 21.5 $Bq\ kg^{-1}$ with a mean value of 11.4 ± 4.6 $Bq\ kg^{-1}$ for $^{232}$Th, from 12.3 to 37.1 $Bq\ kg^{-1}$ with a mean of 23.1 ± 5.9 $Bq\ kg^{-1}$ for $^{238}$U, and from 174 to 320 $Bq\ kg^{-1}$ with a mean of 222 ± 45 $Bq\ kg^{-1}$ for $^{40}$K.

The gamma absorbed dose rates in air were calculated from the specific activities of the $^{232}$Th, $^{238}$U, and $^{40}$K radionuclides according to Equation (1), in a similar manner than that described in details by Tzortzis et al.[2]. The results are presented in Table 3. The calculated total dose rates in air outdoors ranged from 9.6 to 39.4 $nGy\ h^{-1}$ with mean values (A.M. ± S.D.) of 17.3 ± 5.7, 16.7 ± 6.4, 16.5 ± 5.8, 16.4 ± 5.5 and 18.7 ± 5.8 $nGy\ h^{-1}$, for the 5 urban districts shown in Table 1, respectively. The corresponding values in the indoor measurements ranged from 16.7 to 36.0 $nGy\ h^{-1}$ with a mean value of 23.6 ± 5.4 $nGy\ h^{-1}$. In Figure 2, the frequency distribution of the total absorbed dose rates for all the outdoor and indoor measurements is plotted.

Using the calculated indoor and outdoor mean values of the absorbed dose rates in air, the indoor/outdoor ratio is found to be equal 1.4 ± 0.5. This value depends mainly on the activity concentrations of the three naturally occurring radioisotopes ($^{232}$Th, $^{238}$U, and $^{40}$K) in the building and tilling materials used in the house construction. It is expected to be about constant for the urban areas of Cyprus due to the fact that the house construction technique used is very similar everywhere and the building materials are mostly of local origin. It is worth pointing out here, that the calculated indoor/outdoor ratio matches very well with the value of 1.42, which can be derived from worldwide averaged data[1].

The contribution of the $^{40}$K, $^{238}$U and $^{232}$Th radionuclides to the mean total dose rate is found to be almost equal among those three radionuclides (34.7%, 32.5% and 32.8%, respectively) for all the outdoor measurements. This is in accordance with previous results[3]. Furthermore, in the indoor measurements, due to the building materials, a slight enhancement in the relative contribution of the $^{238}$U and $^{40}$K radionuclides to the total dose is observed (36.3%, 38.1% and 25.6%, due to $^{40}$K, $^{238}$U and $^{232}$Th, respectively).

Next, the total effective dose rates to the population were calculated from the outdoor and indoor absorbed dose rate in air, using Equation (2) and assuming that the people spend 20% of their time outdoors and 80% indoors. In the districts where no indoor measurements were performed, the estimated indoor/outdoor ratio of 1.4 for the gamma absorbed rates was used to calculate the indoor effective dose rates for these districts. The calculated total effective dose rates to the population ranged from 78 to 319 $\mu Sv\ y^{-1}$ with a mean value (A.M. ± S.D.) of 138 ± 46 $\mu Sv\ y^{-1}$. As a consequence, the population living in urban areas of Cyprus is subjected to a mean effective dose rate of about 138 $\mu Sv\ y^{-1}$ due to naturally occurring gamma radiation. This important component of the natural radiation has been determined directly, by means of *in situ* high-resolution gamma spectrometry, for the first time and it is representative for the main urban areas of Cyprus.

In the following, an attempt is made to estimate the average total radiation exposure (mean total effective dose) to the inhabitants of the island due to the three most important components of the naturally occurring radioactivity: cosmic rays, terrestrial gamma radiation, and radon ($^{222}$Rn) concentration. According to the UNSCEAR 2000 report[1], the population-weighted average absorbed dose rate from the directly ionising and photon components of the cosmic radiation at sea level corresponds to 31 $nGy\ h^{-1}$, resulting in an effective dose rate of 270 $\mu Sv\ y^{-1}$. From the present study, the mean effective dose rate due to terrestrial gamma radiation is found to be ~138 $\mu Sv\ y^{-1}$, which is obtained from the calculated mean absorbed dose rate of 17.0 $nGy\ h^{-1}$ (outdoors) and 23.6 $nGy\ h^{-1}$ (indoors). This value is by a factor of about 3.5 smaller than the corresponding worldwide average value[1] of 480 $\mu Sv\ y^{-1}$. From the measured radon ($^{222}$Rn) mean concentrations[15] of 19.3 $Bq\ m^{-3}$ (indoors) and 3.9 $Bq\ m^{-3}$ (outdoors), a mean effective dose rate of 527 $\mu Sv\ y^{-1}$ is calculated for the radon component in Cyprus. This value has to be compared with the corresponding worldwide average value[1] of 1095 $\mu Sv\ y^{-1}$. The results are summarized in Table 4. As can be seen, the population in Cyprus is subjected to a total radiation exposure (total effective dose) that exhibits a mean value of 935 $\mu Sv\ y^{-1}$, as far as concern the three components of the natural radiation, which is by a factor of two smaller than the corresponding worldwide average exposure of 1845 $\mu Sv\ y^{-1}$. This implies that the island of Cyprus can be considered as one of the world areas that exhibit very low levels of natural radioactivity. The contribution of each of these three components to the mean total effective dose to the Cyprus population is depicted in Figure 3 in comparison with the corresponding worldwide averaged values.

As a final remark, the results obtained by the present *in situ* high-resolution gamma measurements outdoors are not valid to the small villages situated around the Troodos Mountains, which geologically form a best-preserved ophiolitic complex in the world that is known as "the Troodos ophiolitic complex of Cyprus"[2,3,4]. This region is characterized by igneous rocks, which are very poor in Th and U radioelement concentrations[16]. The activity concentrations of these two radionuclides in the Troodos ophiolithic complex were found to be an order of magnitude lower than the corresponding values observed in the other main formations of Cyprus with sedimentary origin[3,4].

## CONCLUSION

*In situ* high-resolution $\gamma$-ray spectrometry was exploited to determine specific activities and the associated gamma dose rates due to naturally occurring $^{232}$Th, $^{238}$U, and $^{40}$K radioisotopes in 70 outdoor and 20 indoor measurements for the main urban areas of the island of Cyprus. From the measured activities of these radionuclides, gamma absorbed dose rates in air were calculated outdoors and indoors. Hence, the effective dose rate to the population due to terrestrial gamma radiation could be determined accurately. The calculated values ranged from 78 to 319 $\mu Sv\ y^{-1}$ with a mean value of 138 ± 46 $\mu Sv\ y^{-1}$, which is by a factor of 3.5 lower than the value (480 $\mu Sv\ y^{-1}$) derived from worldwide averaged data[1]. Considering all the three components of the naturally occurring radioactivity (terrestrial gamma radiation, cosmic rays, and radon concentration), a mean total effective dose of about 935 $\mu Sv\ y^{-1}$ is calculated for the Cyprus population. This value is by a factor of two lower than the corresponding worldwide mean value[1] (1845 $\mu Sv\ y^{-1}$).

## Acknowledgements

This work was supported by the Cyprus Research Promotion Foundation (Grant No. 45/2001) and, partially, by the University of Cyprus. Special thanks go to Dr. Y. Parpottas for his careful reading of the manuscript.

TABLE CAPTIONS

**Table 1.** District, population[7], and number of outdoor and indoor *in situ* measurements.

**Table 2.** Range and averaged values of the specific activity of naturally occurring $^{232}$Th, $^{238}$U and $^{40}$K radionuclides in outdoor *in situ* measurements performed in the five main urban regions (see Fig .1).

**Table 3.** Range and averaged values of the gamma absorbed dose rate in air from outdoor measurements for the five main urban regions (see Fig .1).

**Table 4.** Mean effective dose to the Cyprus population due to the three components of natural radioactivity compared to worldwide averaged values.

**Table 1.**

| A/A | District | Population (*in thousands*) | Number of outdoor measurements | Number of indoor measurements |
|---|---|---|---|---|
| 1 | Lefkosia | 296.1 | 23 | 20 |
| 2 | Lemesos | 214.8 | 15 | - |
| 3 | Larnaka | 125.2 | 15 | - |
| 4 | Pafos | 71.9 | 11 | - |
| 5 | Ammochostos | 41.2 | 6 | - |
|  | **Total** | 749.2 | 70 | 20 |

**Table 2.**

| A/A | District | Outdoor specific activity ($Bq\ kg^{-1}$) | | | | | |
|---|---|---|---|---|---|---|---|
| | | $^{232}$Th | | $^{238}$U | | $^{40}$K | |
| | | Range | A.M. ± S.D. | Range | A.M. ± S.D. | Range | A.M. ± S.D. |
| 1 | Lefkosia | 5.5 – 25.0 | 10.1 ± 4.4 | 5.4 – 28.8 | 15.7 ± 5.6 | 106 – 388 | 152 ± 58 |
| 2 | Lemesos | 2.9 – 24.1 | 10.4 ± 5.8 | 7.7 – 27.8 | 14.0 ± 5.7 | 99 – 239 | 149 ± 44 |
| 3 | Larnaka | 1.7 – 21.8 | 9.9 ± 5.2 | 6.0 – 24.8 | 13.2 ± 5.0 | 99 – 359 | 159 ± 62 |
| 4 | Pafos | 4.2 – 19.7 | 11.9 ± 4.6 | 7.1 – 21.0 | 11.6 ± 4.1 | 81 – 291 | 146 ± 66 |
| 5 | Ammochostos | 4.7 – 21.5 | 11.6 ± 7.7 | 8.7 – 31.8 | 16.5 ± 8.7 | 102 – 244 | 160 ± 59 |
| | **Total** | 1.7 – 25.0 | 10.6 ± 5.1 | 5.4 – 31.8 | 14.2 ± 5.7 | 81 – 388 | 153 ± 56 |

**Table 3.**

| A/A | District | Total outdoor dose ($nGy\ h^{-1}$) | | |
|---|---|---|---|---|
| | | Min | Max | A.M. ± S.D. |
| 1 | Lefkosia | 11.8 | 39.4 | 17.3 ± 5.7 |
| 2 | Lemesos | 10.3 | 32.4 | 16.7 ± 6.4 |
| 3 | Larnaka | 10.0 | 35.0 | 16.5 ± 5.8 |
| 4 | Pafos | 9.6 | 25.9 | 16.4 ± 5.5 |
| 5 | Ammochostos | 11.0 | 26.3 | 18.7 ± 5.8 |
| | Total | 9.6 | 39.4 | 17.0 ± 5.7 |

**Table 4.**

| A/A | Type of natural radioactivity | Mean total effective dose ($\mu Sv\ y^{-1}$) | |
|---|---|---|---|
| | | Cyprus | World average [1] |
| 1 | Gamma radiation | 138 | 480 |
| 2 | Cosmic rays (sea level) | 270[1] | 270 |
| 3 | Radon ($^{222}$Rn) | 527[15] | 1095 |
| | Total | 935 | 1845 |

FIGURE CAPTIONS

**Figure 1**. Simplified map of Cyprus showing the 5 main districts studied. The locations of the main towns (indicated by the solid circle points), in and around of which the *in situ* measurements were performed, are also depicted.

**Figure 2.** Frequency distribution of the total absorbed dose rate in air due to gamma radiation for all the *in situ* measurements.

**Figure 3.** The contribution of each of the three components of the naturally occurring radioactivity to the mean total effective dose to the Cyprus population compared to the corresponding worldwide averaged values.

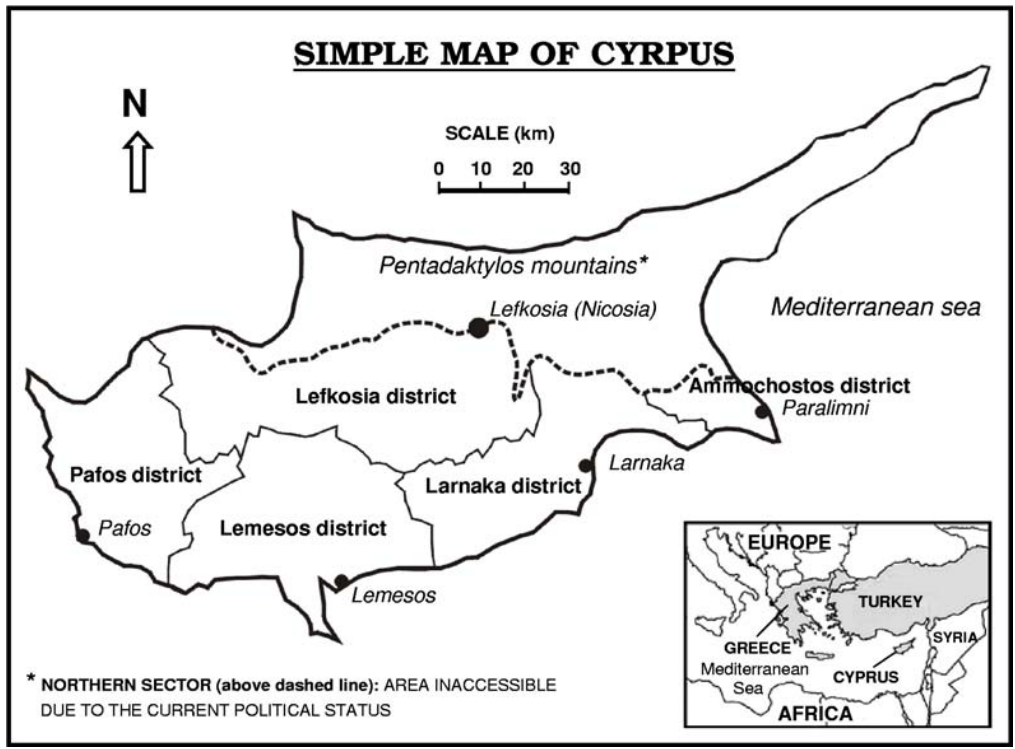

**Figure 1**.

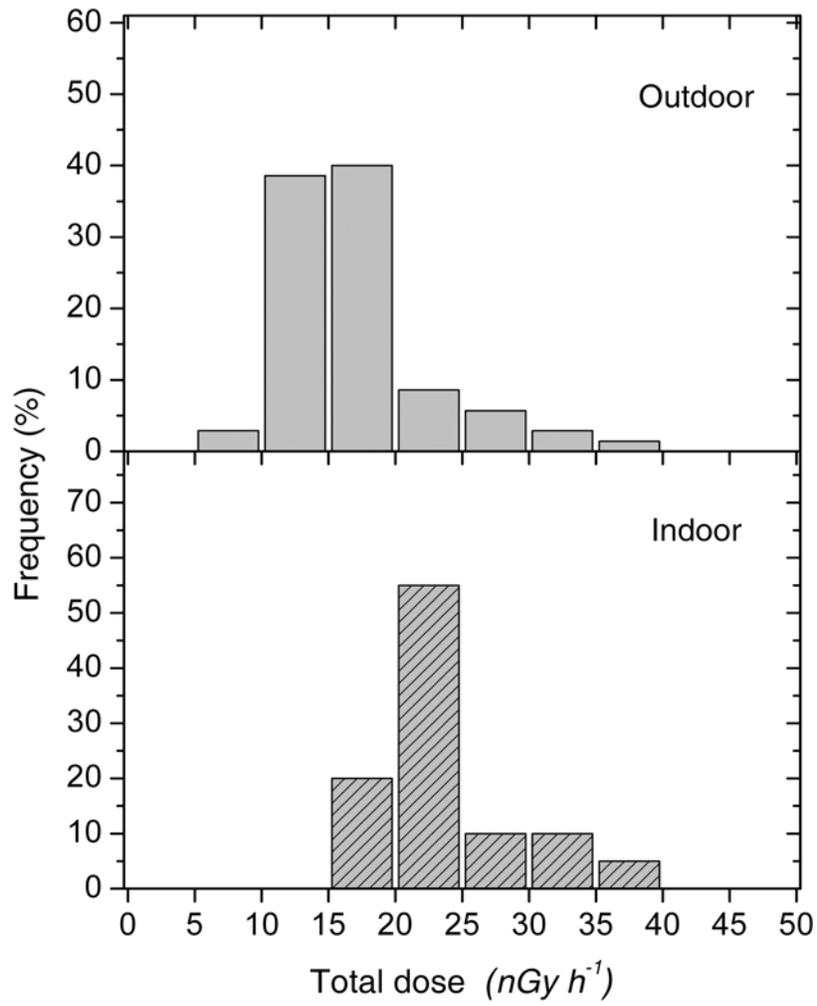

**Figure 2**

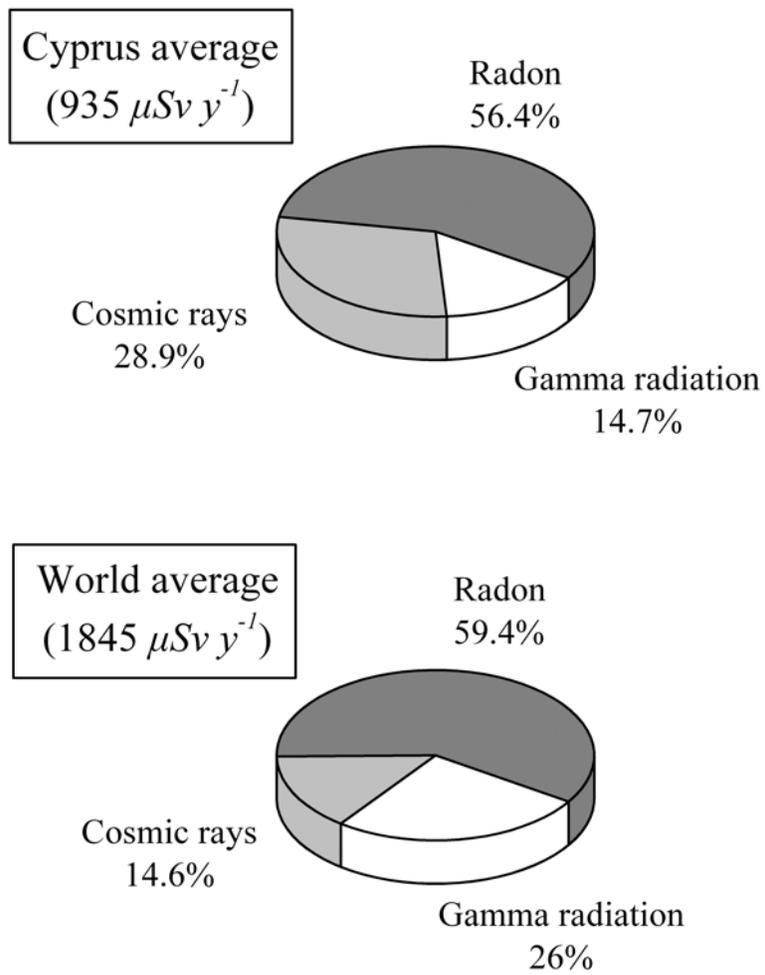

**Figure 3**